\documentclass[10pt,3p]{elsarticle}
\usepackage{graphicx}
\usepackage{float}
\usepackage{subfig}
\usepackage{setspace}
\usepackage{color}
\usepackage{multirow}
\usepackage{amsmath}

\begin{document}




\title{Control of a Backward-Facing Step flow through vortex pairing and phase locking}


\author[1,2]{Thomas Duriez}
\author[1]{Jean-Luc Aider}
\author[1]{Jose Eduardo Wesfreid}
\author[2]{Guillermo Artana}

\address[1]{Laboratoire PMMH, UMR CNRS 7636, ESPCI ParisTech, Universit\a'e Pierre et Marie Curie, Universit\a'e Paris Diderot}
\address[2]{Laboratorio de Fluidodinamica, Facultad de Ingeneria, Universidad de Buenos Aires}

\begin{abstract}
Many experimental and numerical studies report a large reduction of the recirculation bubble in Backward-Facing Step flows or airfoils in stall situation when excited at the natural shedding frequency $f_0$. Through a simple experiment using Dielectric Barrier Discharge actuator, we find a different result.  For a given Reynolds number, the frequency of the perturbation is varied for a fixed duty-cycle dc = 27\%. Through phase-averaging of Particle Image Velocimetry measurements, we show that the actuation creates a forced vortex which interacts with the natural shedding with a different phase velocity than the unforced one. The largest reduction of the recirculation bubble (-35\%) is obtained in a very narrow frequency range around $0.73 f_0$ where early vortex pairing occurs between forced and unforced vortices. Phase averaging shows that in this case, the actuation clearly forces the vortex pairing in the shear layer. On the contrary, when the forcing frequency is higher, the shear layer behaves like an amplifier synchronized on the forced frequency, leading to a constant 10\% reduction of the recirculation bubble. 
\end{abstract}


\maketitle






\section{Introduction}
\label{intro}
It is well-known that the Backward-Facing Step (BFS) flow is very sensitive to upstream flow conditions and exhibits multiple natural characteristic frequencies which are associated to various phenomena like Kelvin-Helmholtz instability, flapping of the shear layer or oscillation of the recirculation bubble~\cite{jla07} and global 3D instabilities \cite{Beaudoin_EJM,Barkley_step02}. In the following, we focus on the control of a massively separated flow downstream a backward-facing step using temporal actuation just upstream of the BFS edge.

Indeed there are experimental and numerical evidences \cite{chun-sung-96,Wengle2001,Henning2007,LeQuere10,Dandois07} of a large reduction of the recirculation bubble (up to -40\%) when tuning the excitation frequency $f_e$ around the natural frequency of  the most amplified Kelvin-Helmholtz frequency $f_0$ of the shear layer, corresponding to a Strouhal number $St_\theta = \frac{f_0\theta}{U_0} = 0.0175$ based on $\theta$  the momentum layer thickness measured near the separation point. 

Most of the previous studies use pulsed jets interacting with the incoming boundary layer with a $45^{o}$ or $90^{o}$ angle \cite{chun-sung-96,Wengle2001,Pastoor2008,LeQuere10}. Nevertheless, one can think that the way the perturbation is injected into the shear layer will influence its growth and amplification. 
In  this study, a blowing tangential to the wall, homogenous along the spanwise direction, is injected just upstream the BFS edge. For this purpose, we use a classic surface Dielectric-barrier discharge (DBD) plasma actuator\cite{Moreau07}. This perturbation is the best way to force the shear layer without injecting  3D perturbations in the same time. Unlike other studies, the perturbation is streamwise and the perturbation ratio is not small compared to one, in order to induce non-linear effects.

\section{Experimental Set-Up}
\label{sec:1}
\subsection{Wind tunnel}
\label{subsec:1}

\begin{figure}
\begin{center}
\includegraphics[width=0.8\textwidth]{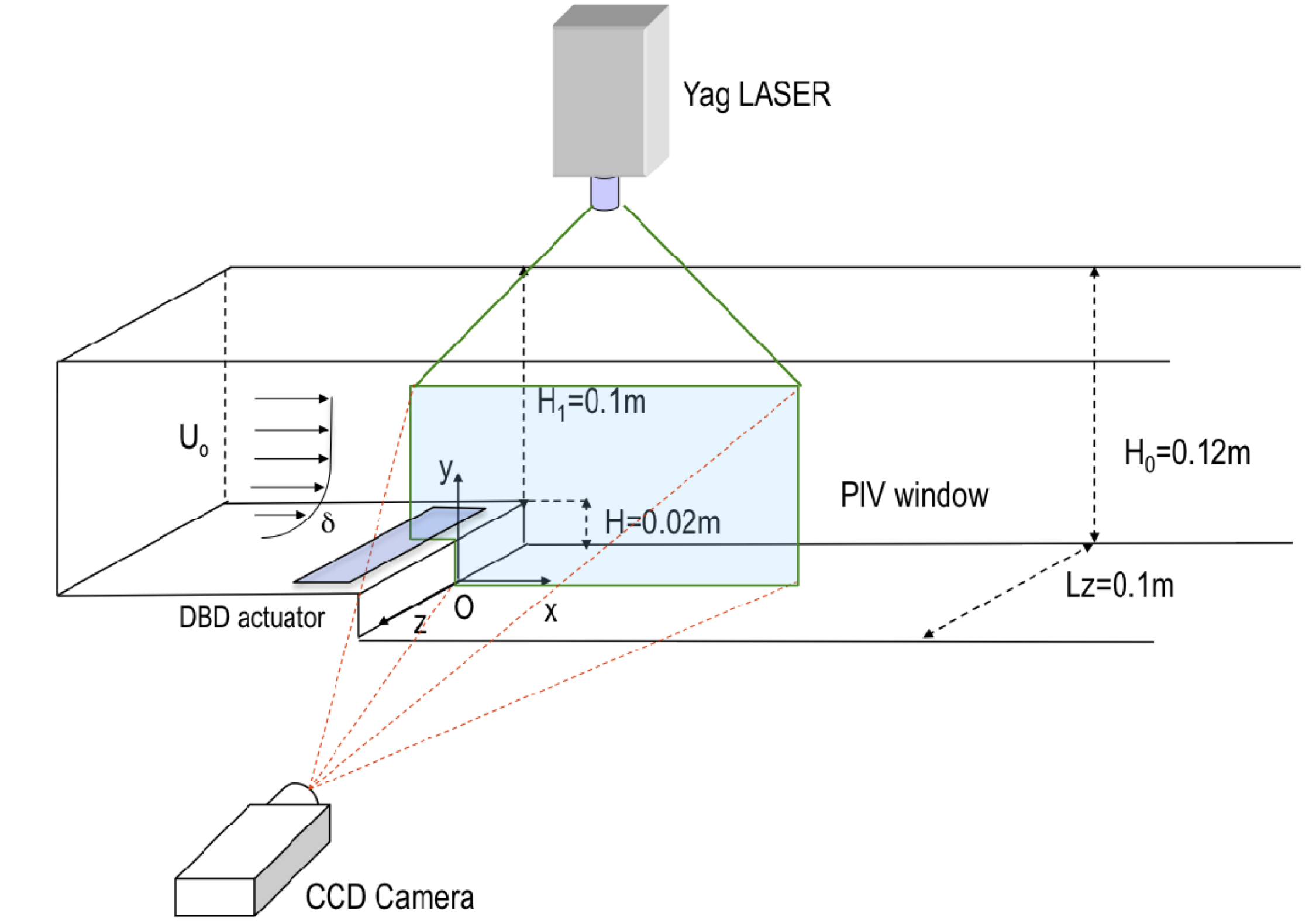}
\end{center}
\caption{\label{fig:setup} Test section geometry and position of the actuator and PIV setup.  }
\end{figure}

The experimental setup consists in a low-speed wind tunnel already described in Beaudoin et al\cite{Beaudoin_JFS06}. The test section has a 0.1$\times$0.1m$^2$ square cross-section. The step-height is $H = 0.2$ $m$ (Fig.~\ref{fig:setup}) and thus the expansion ration is $E=\frac{H}{H_0}=0.17$, with $H_0$ the total height of the channel.  
All the results discussed in the following have been obtained for a given Reynolds number $Re_H = \frac{U_0H}{\nu} = 4400$, corresponding to a free-stream velocity $U_0 = 3.3 m.s^{-1}$. In this case, the most amplified or natural shedding frequency $f_0 \approx 69Hz$ is measured through time-resolved visualizations using a fast camera (1kHz) and smoke injection. From the initial unstable velocity profile measured at step edge, we measure  the momentum layer thickness $\theta = 0.8mm$. The corresponding Strouhal number $St_\theta = 0.0167$ close to the one found in the previous studies as mentioned in the introduction. The downstream variation of the shear layer and  its  streamwise spreading rate is measured $\frac{d\theta}{dx} = 0.004$ at the step edge and $\frac{d\theta}{dx} = 0.05$  for $x>0,8H$. This is in good agreement with the values obtained in other mixing layers\cite{Ho-Huerre1984}. 

\subsection{DBD Plasma discharge}
\label{subsec:2}
\begin{figure}
\begin{center}
\includegraphics[width=0.8\textwidth]{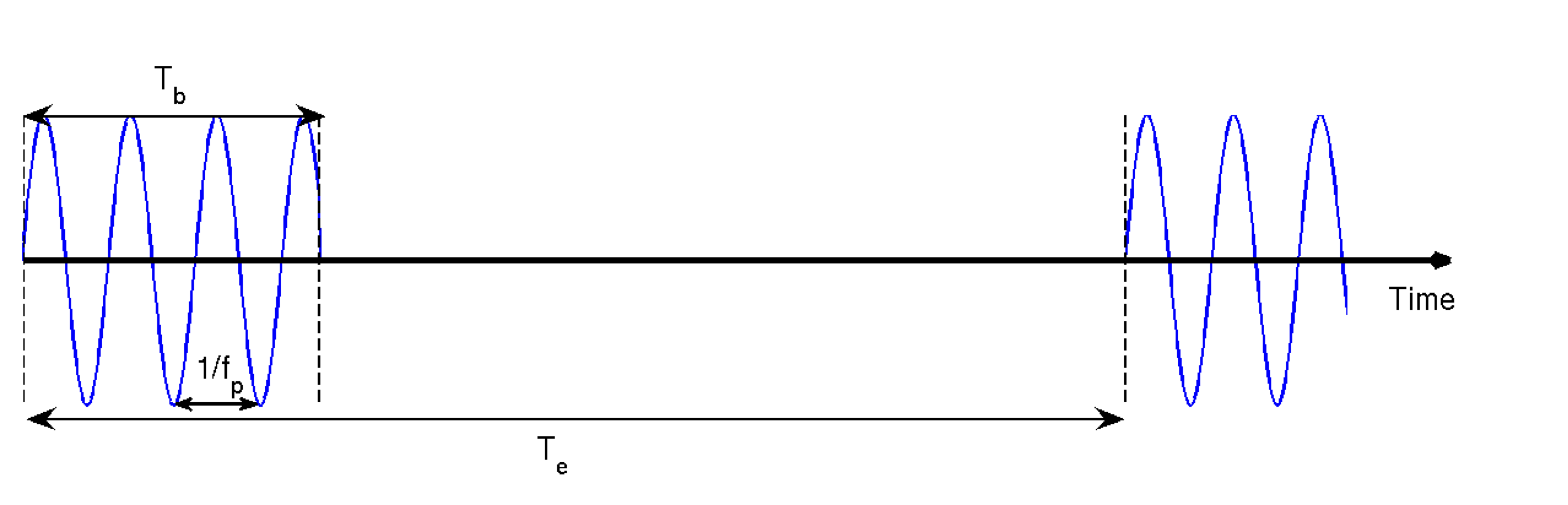}
\end{center}
\caption{\label{fig:tdia} Time diagram of the actuation.}
\end{figure}

The plasma actuation is obtained by applying a high frequency ($f_p=3.9$kHz) AC voltage (10 KV) between two electrodes separated by an insulating mylar dielectric barrier to produce an electrical discharge. It  generates a ionic wind parallel to the wall and in the direction of the flow with a jet velocity $u_j\approx 4 m.s^{-1}$ (measured by PIV).  A square waveform biases the voltage to create the pulsed actuation.  $T_e=1/f_e$ is the excitation time, $T_b=0.27T_e$ is the blowing time while the actuator is on and $1/f_p$ is the period of the AC voltage signal of the electric discharge.The duty cycle is fixed at $\frac{T_b}{T_e}=27\%$  (figure~\ref{fig:tdia}). The frequency of the waveform $f_e$, which is the control parameter, varies between 20 to 150 Hz corresponding to a non-dimensional frequency  $F+ =  \frac{f_e}{f_0}$ ranging between $ 0.3$ and $ 2.2$. The discharge is placed $5mm$ upstream of the separation.

\subsection{PIV system and phase reconstruction}
\label{subsec:3}
The natural and the perturbed flows are studied using a standard Particle Image Velocimetry (PIV) setup with a 4Hz YaG laser synchronized with a double frame PIV camera. For every configurations, 720 double-frame pictures are recorded so that the mean velocity vector fields are well defined and converged. The PIV setup is also synchronized on the waveform signal applied on the actuator to enable phase averaging and analysis of the fluidic interactions during an actuation cycle. The actual acquisition frequency is determined in order to avoid stroboscopic effect with the forcing frequency.

\section{Results}
\label{sec:2}
\subsection{Natural flow}
\label{subsec:4}
\begin{figure}
\begin{center}
\includegraphics[width=0.8\textwidth]{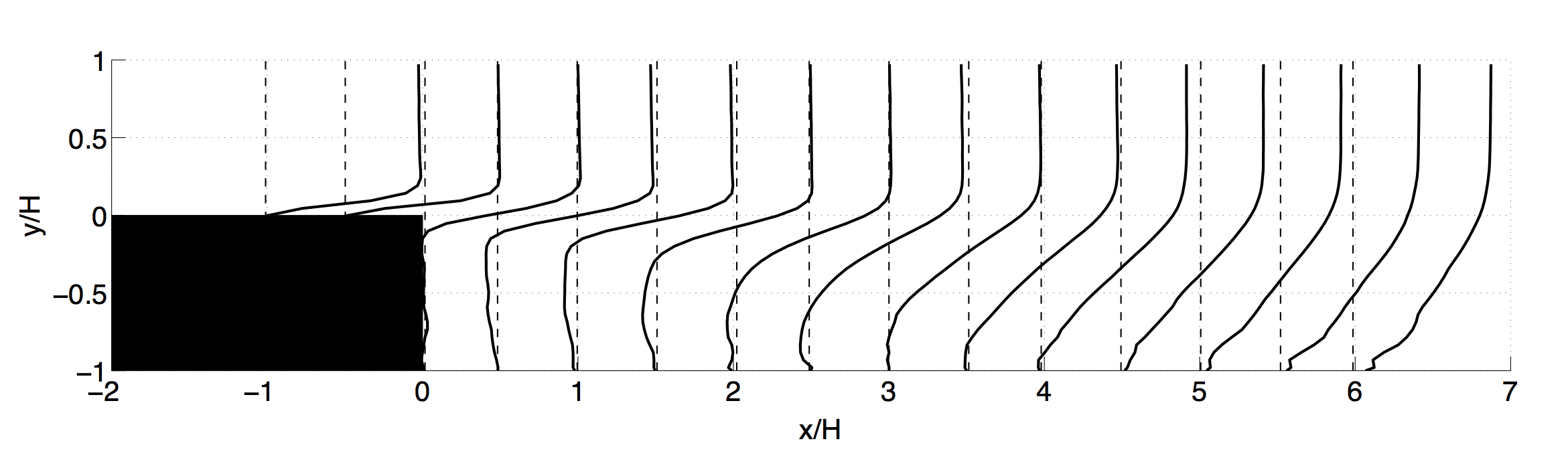} 
\caption{\label{fig:PIVmoyens_Insta} a) Time-Averaged streamwise velocity profiles of the unforced flow at $Re_H = 4400$.}
\end{center}
\end{figure}

The natural flow is illustrated on Fig.~\ref{fig:PIVmoyens_Insta} showing time averaged profiles of the streamwise velocity, which, with contours of time-averaged streamwise velocity $<U(x,y)>_t$ are used to measure the length of the recirculation bubble $X_r$ which will be modified by the actuation. In the following, $X_{r0}$ is the length of the unforced recirculation bubble. In our case, $X_{r0} = 4.2H$. 

\subsection{Actuated flow}
\label{subsec:5}
\begin{figure}
\begin{center}
\includegraphics[width=0.8\textwidth]{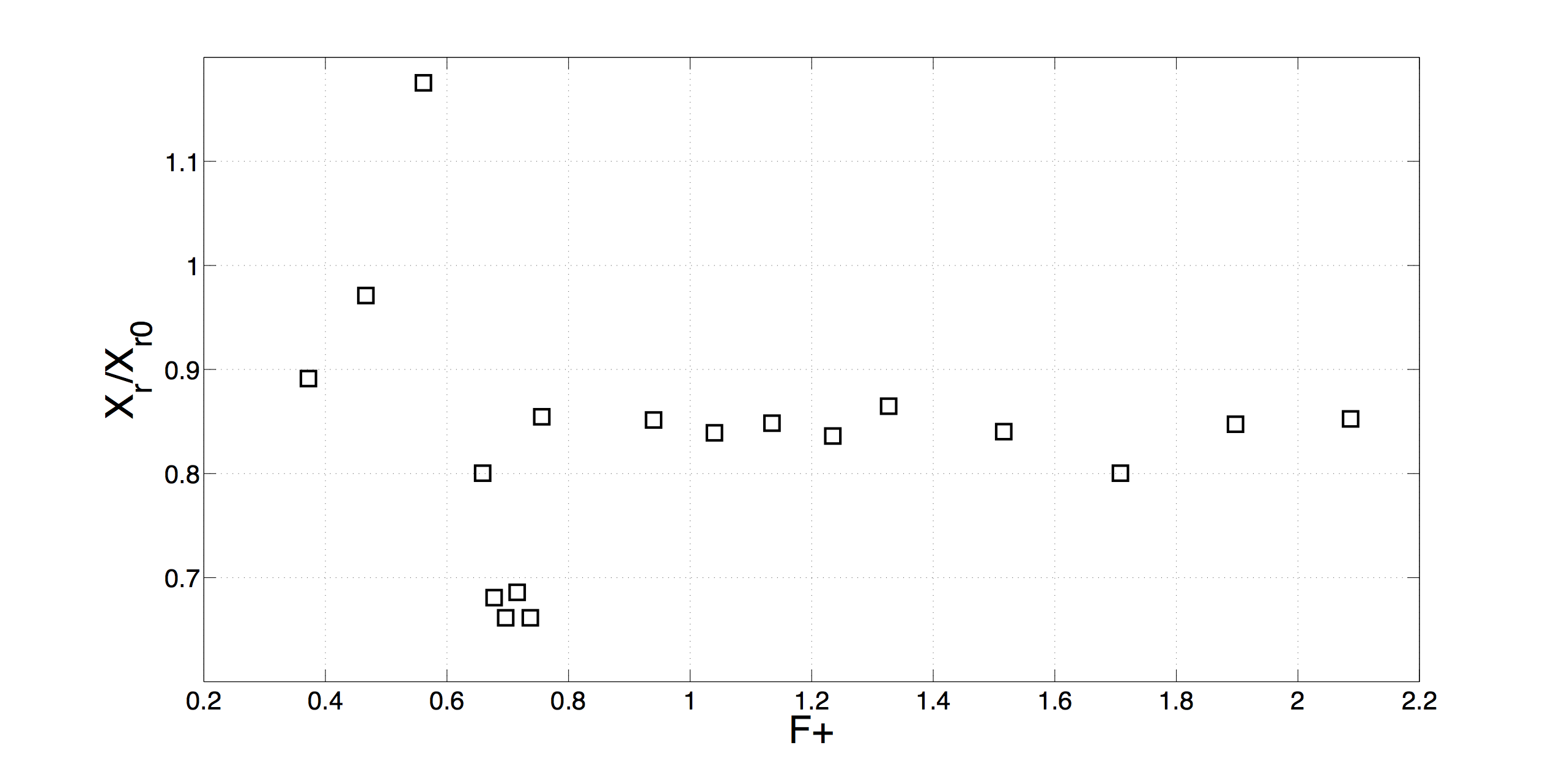}
\caption{\label{fig:LrF} Evolution of $\frac{X_r}{X_{r0}}$ as a function of the forcing frequency $F+$.}	
\end{center}
\end{figure}

The first evidence of the influence of the pulsed actuation is the modification of  $X_r$. As illustrated on Fig.~\ref{fig:LrF}, a large reduction of the recirculation length (-34\%) is found when the actuation frequency reaches $F+= 0.73$. This is different from previous studies \cite{chun-sung-96,LeQuere10} where the maximum reduction is obtained when the excitation matches the natural shedding frequency ($F+\approx 1$) in a relatively large frequency band. In this case this effect occurs in a very narrow frequency band and the optimal excitation frequency is lower than the natural shedding frequency. It is important to notice that the 0.73 coefficient corresponds exactly to the percentage of the time when the actuation is off during one cycle ($\frac{T_b}{T_e} = 0.27$). More precisely, the recirculation bubble is minimum when the time when the actuation is off corresponds to the natural shedding period $T_e - T_b = T_0$.  

For lower forcing frequencies the recirculation length is reduced (except for $F+ = 0.57$ which leads to a +18\% increase of the recirculation bubble). For larger frequencies a nearly constant -10\% reduction of $X_r$ is observed with a local minimum for $F+= 1.7$.

\subsection{Actuated flow dynamics}
\label{subsec:6}
\begin{figure}
\begin{center}
\includegraphics[width=0.7\textwidth]{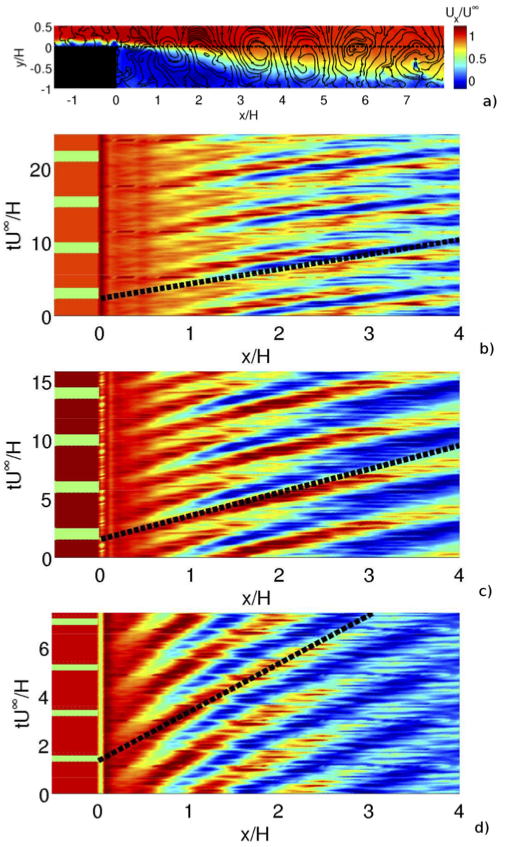}
\caption{\label{fig:spatio} Spatio-temporal diagrams of a horizontal line ($y=0$) (a) in the phase-averaged streamwise velocity fields  for $F+= 0.43$ (b), $F+= 0.73$ (c) and $F+= 1.5$ (d).}
\end{center}
\end{figure}

To try to understand why the most efficient frequency is $F+= 0.73$, it is necessary to have more insights into the transient phenomena happening during one excitation cycle. In this purpose the time sequence during one cycle is reconstructed using phase averaging. The cycle is  decomposed into 36 events taken from the PIV time series. It shows clearly the shedding of KH vortices (Fig.~\ref{fig:spatio}a). From these velocity fields it appears that the dynamics of the KH vortices is strongly affected by the actuation.

To clarify the underlying mechanism, we plot on Fig.~\ref{fig:spatio} the spatio-temporal diagrams of a horizontal line ($y=0$ and $x = 0$ to $6H$) in the phase-averaged streamwise velocity fields  for three different actuation frequencies. On the left of the diagrams, the actuator appears as green stripes when on and red stripes when off. $x/H=0$ corresponds to the step edge.The red and blue oblique stripes correspond to the traces of the vortices convected in the shear layer. Their slopes correspond to the phase velocity $u_\phi$ of the given vortex. The dark dotted lines shows the slope of constant phase velocity $u_\phi$ equal to $U_0/2$. Different phenomena are observed in the region 0$ \leq \frac{x}{H} \leq$ 2,  depending on the forcing frequency. 

\begin{figure}
\begin{center}
\includegraphics[width=0.8\textwidth]{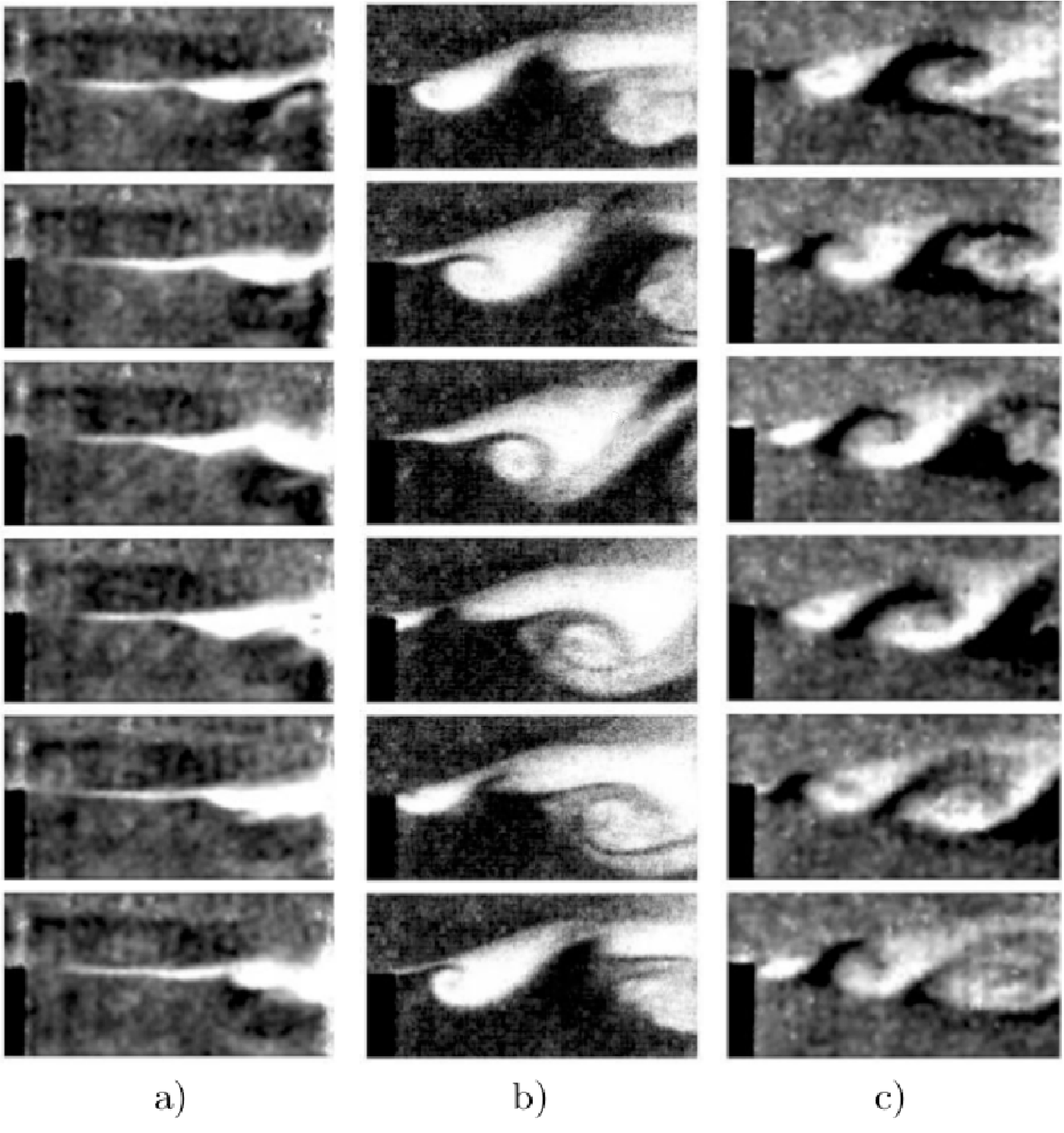}
\caption{\label{fig:pairing} Visualisation of unforced a) and forced flows b) and c).
On figure b) the smaller forced vortex catches-up the unforced one and initiate a vortex pairing, which leads to a large scale vortical structure ($F+ \approx 0.73$). On Figure c) the shear layer is synchronized by the excitation ($F+ >0.73$).}
\end{center}
\end{figure}

\begin{itemize}
\item When $F+<0.73$  (Fig.~\ref{fig:spatio}a), several vortices are shed in a single cycle. In addition to the \emph{forced} one, one or several \emph{natural} vortices are shed at random phase shifts which leave traces in the spatio-temporal diagrams.
\item When $F+\approx 0.73$  (Fig.~\ref{fig:spatio}b). Two vortices are shed in a single cycle. A \emph{forced} vortex is shed during the excitation(when the actuator is on). A \emph{unforced} one is shed while the actuator is off. The shedding of the \emph{unforced} vortex has a constant phase shift compared to the shedding of the \emph{forced} vortex. Moreover, the forced vortex having a much higher phase velocity than the unforced one, it catches up the former, leading to vortex pairing and to a large reduction of the recirculation length. The difference in phase velocity can be accounted on the large jet velocity ($\frac{u_j}{U_0} \approx 1.2$).
\item When $F+ >0.73$ (Fig.~\ref{fig:spatio}c). A single vortex is shed at the exact forcing frequency. The shear layer behaves like a forced oscillator following perfectly the actuation frequency.
\end{itemize}

These results are confirmed by visualizations obtained by smoke injection and fast-camera recordings (figure~\ref{fig:pairing}). The vortex pairing can be observed on Fig.~\ref{fig:pairing}b.

\begin{figure}
\begin{center}
\includegraphics[width=0.8\textwidth]{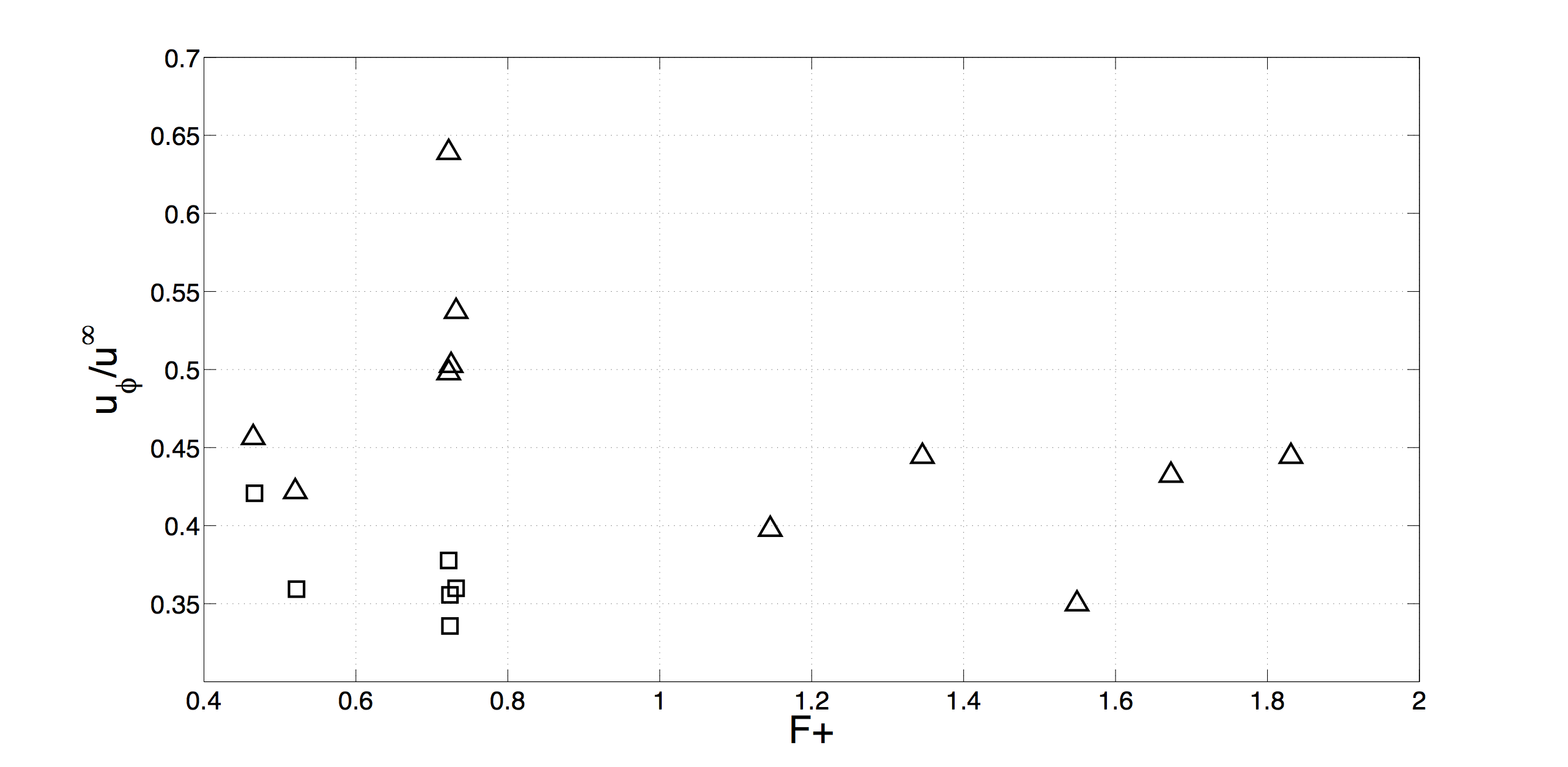}
\caption{\label{fig:phase_velocity} Evolution of the phase velocity $u_\phi$ of the forced (open squares) and unforced ($\triangle$) vortices as a function of the forcing frequency $F_+$.}
\end{center}
\end{figure}

On Fig.~\ref{fig:phase_velocity}, the phase velocity of unforced and forced vortices shed from the BFS is plotted as a function of the forcing frequency. For $F+ > 0.73$ the only vortex shed during one cycle has a phase velocity around 0.4. For lower Strouhal numbers \emph{forced} vortices have a slightly higher phase velocity (up to ${\Delta}u_\phi/U_0=0.07$) than \emph{unforced} ones. This difference becomes much higher (between ${\Delta}u_\phi/U_0=0.15$ to ${\Delta}u_\phi/U_0=0.25$) at the optimal forcing frequency. 

\section{Discussion}
\label{sec:3}
This difference can be explained using vortex-vortex interactions. For higher Strouhal numbers only one type of vortex is shed very regularly  so that the flow fields induced by both preceding and following vortices are always the same. The effect induced by the preceding vortex is counterbalanced by the following one which have the same vorticity and position. This leads to a phase velocity with small variation compared to the natural flow. 

On the contrary, when $T_e - T_b > T_0$ unforced and forced vortices are not shed in the same conditions. Unforced vortex are always shed after a time $T_0$ from the previous forced vortex. Forced vortex are shed with a time which depends on $f_e$. When $T_e - T_b = T_0$ the delay between the forced vortex and the preceding unforced one is minimum, while the delay with the next vortex is maximum ($T_0$), creating a strong interaction between those two vortices. The forced vortex is most likely shed with a slightly higher advection velocity due to the blowing during $T_b$. Moreover it appears that the respective size and position of the vortices provoke the early rising of the forced vortex in high advection velocity area and the fall of the unforced one where advection is slower (fig.~\ref{fig:pairing}b) leading to the highest difference in phase velocity detected in the figure~\ref{fig:phase_velocity}. The later the forced vortex is shed, the weaker is the interaction with the preceding vortex, and the more counter-balanced it is by the following vortex, leading to a lower difference in the phase velocity. Thus the pairing, if any, is achieved farther from the separation point. The main difference between this study and earlier work pointing a minimum of the recirculating happening at a Strouhal number close to the most amplified by the shear layer lies in the amplitude of the actuation. It has been pointed out that sub-harmonic phenomenons in the shear layer need high actuation level. The fact that no multi-scale vortex pattern is found like pointed out by earlier experimental and numerical studies may lie in the fact that the actuation is upstream of the separation and its main effect is not pressure driven and thus acting on the whole shear layer.

\section{Conclusion}
\label{conclu}
In this study, we find that a forcing frequency $F+= 0.73$ leads to a large reduction of the recirculation bubble. The maximum reduction is associated to an early pairing of successive shed vortices with a high difference in their phase velocity. The optimal frequency is obtained when $T_e - T_b = T_0$ i.e. when the natural shedding period corresponds to the time when the actuator is off. This configuration is also the one where phase velocity difference between the forced vortex and the unforced one is the greatest. It allows an early vortex pairing between forced and unforced vortice, enhancing the mixing in the shear layer and thus decreasing the size of the recirculation bubble. For higher forcing frequencies the system cannot shed another vortex when the actuator is off. In this case only the forced vortex is shed during one cycle at the forcing frequency. For lower frequencies the shedding is irregular as we can follow similar vortices shed at different phase instants. In addition the distance between the forced and unforced vortices is necessarily higher in this case than in the optimal case. This is different from past studies as we are actuating with high relative velocity and with a jet that has an effect on the phase velocity due to its parallelism to the main flow. This can be a condition to achieve early vortex pairing.






\bibliographystyle{unsrt}	

\bibliography{citations_marche2}

\begin{thebibliography}{10}

\bibitem{jla07}
J.-L. Aider, A.~Danet, and M.~Lesieur.
\newblock Large-eddy simulation applied to study the influence of upstream
  conditions on the time-dependant and averaged characteristics of a
  backward-facing step flow.
\newblock {\em J. Turbulence}, 8, 2007.

\bibitem{Beaudoin_EJM}
J.-F. Beaudoin, O.~Cadot, J.-L. Aider, and J.~E. Wesfreid.
\newblock Three-dimensional stationary flow over a backward-facing step.
\newblock {\em Eur. J. Mech. B/Fluids}, 23(1):147--155, 2004.

\bibitem{Barkley_step02}
D.~Barkley, M.~G.~M. Gomes, and R.~D. HENDERSON.
\newblock Three-dimensional instability in flow over a backward-facing step.
\newblock {\em J. Fluid Mech.}, 473:167--190, 2002.

\bibitem{chun-sung-96}
K.~B. Chun and H.~J. Sung.
\newblock Control of turbulent separated flow over a backward-facing step by
  local forcing.
\newblock {\em Exp. Fluids}, 21:417--426, 1996.

\bibitem{Wengle2001}
H.~Wengle, A.~Huppertz, G.~B\"{a}rwolff, and G.~Janke.
\newblock The manipulated transitional backward-facing step flow: an
  experimental and direct numerical investigation.
\newblock {\em Eur. J. Mech. B/Fluids}, 20:25--46, 2001.

\bibitem{Henning2007}
Lars Henning and Rudibert King.
\newblock {Robust Multivariable Closed-Loop Control of a Turbulent
  Backward-Facing Step Flow}.
\newblock {\em Journal of Aircraft}, 44(1):201--208, January 2007.

\bibitem{LeQuere10}
Z.~Mehrez, M.~Bouterra, A.~El~Cafsi, A.~Belghith, and P.~{Le Quere}.
\newblock Mass transfer control of a backward-facing step flow by local
  forcing-effect of reynolds number.
\newblock {\em Thermal Science}, 00(4):47, 2010 Online-First.

\bibitem{Dandois07}
J.~Dandois, E.~Garnier, and P.~Sagaut.
\newblock Numerical simulation of active separation control by a synthetic jet.
\newblock {\em J. Fluid Mech.}, 574:25--58, 2007.

\bibitem{Pastoor2008}
M.~Pastoor, L.~Henning, B.~R. Noack, R.~King, and G.~Tadmor.
\newblock Feedback shear layer control for bluff body drag reduction.
\newblock {\em J. Fluid Mech.}, 608:161--196, 2008.

\bibitem{Moreau07}
E.~Moreau.
\newblock Airflow control by non-thermal plasma actuators.
\newblock {\em J. Phys. D Appl. Phys.}, 40(3):605, 2007.

\bibitem{Beaudoin_JFS06}
J.-F. Beaudoin, O.~Cadot, J.-L. Aider, and J.~E. Wesfreid.
\newblock Bluff-body drag reduction by extremum-seeking control.
\newblock {\em J. Fluids Struc.}, 22(6-7):973 -- 978, 2006.

\bibitem{Ho-Huerre1984}
C.M. Ho and P.~Huerre.
\newblock Perturbed free shear layers.
\newblock {\em Annu. Rev. Fluid Mech.}, 16:365--424, 1984.

\end{thebibliography}







\end{document}